\documentclass[5p]{elsarticle}

\usepackage{multirow,url,graphicx,siunitx,textcomp,helvet,array,makecell,hhline,lineno,hyperref,epstopdf,chemformula}
\newcolumntype{M}[1]{>{\centering\arraybackslash}m{#1}}
\DeclareSIUnit\px{px}
\modulolinenumbers[5]
\usepackage{amssymb}

\usepackage{tikz}
\usepackage{pgfplots}
\pgfplotsset{width=0.375\textwidth,compat=1.9}
\pgfplotsset{compat=newest}
\usepgfplotslibrary{units}
\journal{Measurement}

\begin{document}

\begin{frontmatter}
\title{Thermometry of intermediate level nuclear waste containers in multiple environmental conditions}
\author{J Norman\fnref{fn1}}
\author{A Sposito, J L McMillan, W Bond, M Hayes, R Simpson, G Sutton, V Panicker and G Machin}
\address{National Physical Laboratory, Hampton Road, Teddington, TW11 0LW, UK}
\author{J Jowsey and A Adamska}	
\address{Sellafield Site, Seascale, Cumbria, CA20 1PG}
\fntext[fn1]{Corresponding author \texttt{james.norman@npl.co.uk}}

\begin{abstract}
Intermediate level nuclear waste must be stored until it is safe for permanent disposal. Temperature monitoring of waste packages is important to the nuclear decommissioning industry to support management of each package. Phosphor thermometry and thermal imaging have been used to monitor the temperature of intermediate level waste containers within the expected range of environmental storage conditions at the Sellafield Ltd site: temperatures from \SIrange{10}{25}{\celsius} and relative humidities from \SI{60}{\percent}rh to \SI{90}{\percent}rh. The feasibility of determining internal temperature from external surface temperature measurement in the required range of environmental conditions has been demonstrated.
\end{abstract}

\end{frontmatter}


\section{\bf Scope of Work} \label{sec:scope}
Nuclear fission is a major part of the energy infrastructure of the UK. However, the decommissioning of nuclear facilities requires the safe and sustainable storage of spent fuel and other radioactive by-products. One form of this waste, intermediate level waste (ILW), mostly comprises nuclear reactor fuel element cladding and components, and radioactive liquid effluent sludges, both of which are immobilised in grout. Other wastes, such as graphite and various scrap metal components, are stored in ILW waste containers without grout encapsulation \cite{ref:nda_ilw_waste}. Typically at the Sellafield Ltd site ILW is stored in cylindrical steel containers and described as waste packages when filled. The container has a pair of dewatering tubes, a sintered gauze layer above the waste, and a meshed vent on the container lid.

Recent laboratory-based measurements of an ILW container demonstrated a correlation between the temperature measured from internal contact thermometers and the external vent radiance temperature \cite{ref:ilw_paper_2018}. To understand the challenges applying these laboratory measurements to the ventilated engineered stores used, this experimental design was replicated in an environmental chamber to simulate the varying temperatures  and high humidity typically experienced. The two methods employed by the temperature group at the National Physical Laboratory (NPL) to determine the vent temperature were: thermal imaging and phosphor thermometry \cite{ref:phosphor_thermometer_reference}.


\section{\bf Experimental Setup} \label{sec:experimental_setup}
This section details the ILW container instrumentation setup and the testing environment. The measurements of the container vent temperature using a thermal imager and phosphor thermometer are described.

\subsection{Container configuration}\label{subsec:measurements_container}
\begin{figure}[t]
\centering
\includegraphics[width=0.45\textwidth,keepaspectratio]{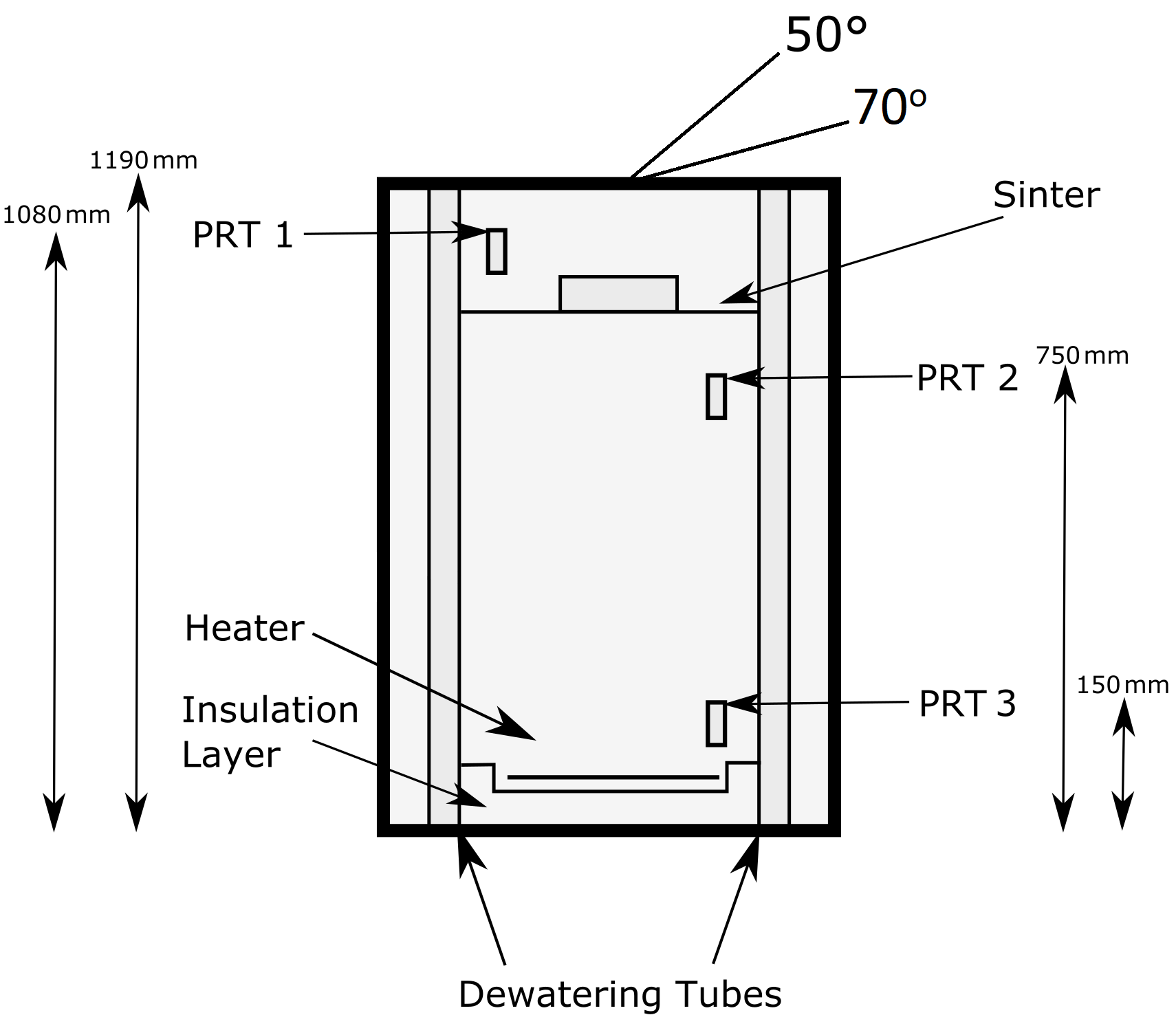}
\caption{A schematic of the internal instrumentation of the ILW container. Distances were measured from the bottom of the container to the top of the PRT. The angles of observation depict the positions used for the thermal imager.}
\label{fig:drumSchematic}
\end{figure}

A schematic of the ILW container can be seen in Fig.~\ref{fig:drumSchematic}, the experimental conditions are identical to those in \cite{ref:ilw_paper_2018}. To supply heat to the container, two heaters were placed at the bottom of the container above the insulation layer, comprising both wooden blocks and multi-layer insulation. The heaters were each connected to a benchtop controller that regulated the power to the heaters, to obtain a stable temperature set-point.

\begin{figure}[t]
\centering
\includegraphics[width=0.45\textwidth,keepaspectratio]{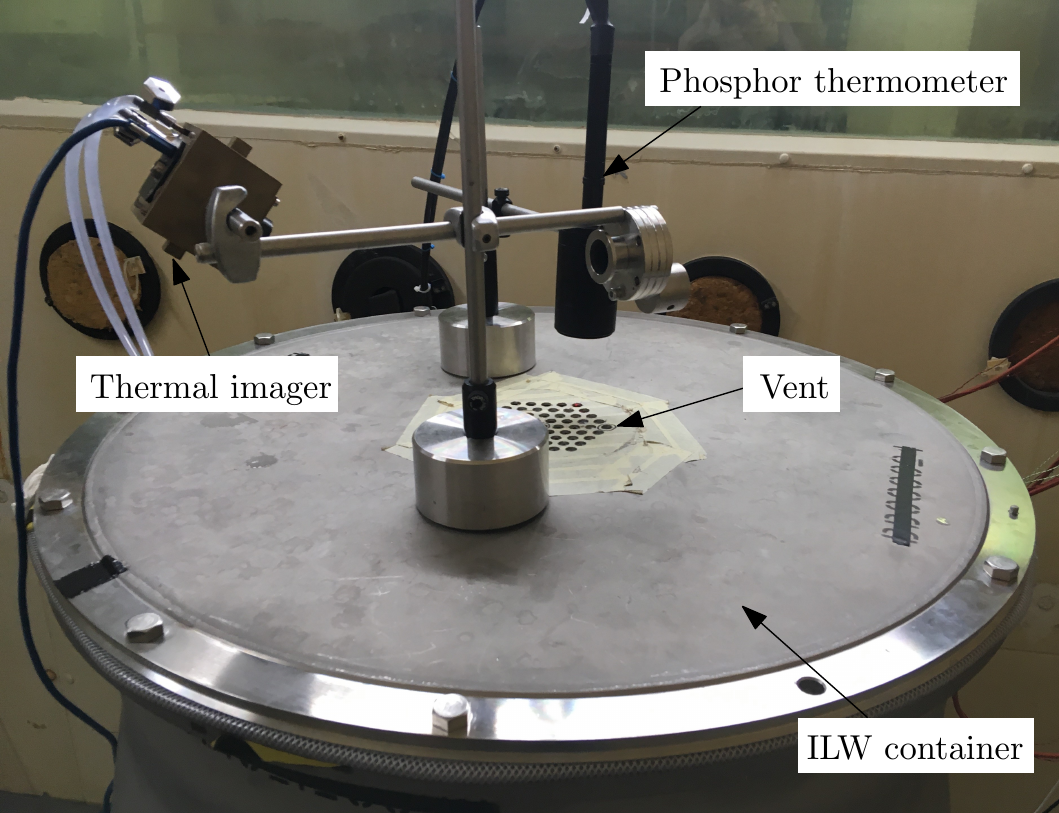}
\caption{The thermal imager and phosphor thermometer in position above the vent of the ILW container lid. The ILW container is in an environmentally controlled chamber~\cite{ref:environ_chamber}.}
\label{fig:setup_pic}
\end{figure}

The ILW container was set up (see Fig.~\ref{fig:setup_pic}) within a large environmental chamber situated at the Office for Product Safety and Standards~\cite{ref:environ_chamber}. This facility permits the control of temperature from \SIrange{-25}{70}{\celsius} and humidity up to \SI{95}{\percent}rh and is regulated using a vented air flow system --- the system results in significant movement of air through the chamber. 

Three class A thin film Pt100 platinum resistance thermometers (PRTs) \cite{ref:iso751} were each potted in a \SI{40}{\milli\metre} long cylinder (\SI{4}{\milli\metre} in diameter). These PRTs were fixed to the internal dewatering tubes (gauzed and hollow tubes, seen in Fig.~\ref{fig:drumSchematic}) within the ILW container to evaluate the internal bulk temperature. 

A range of parameters were investigated in this measurement campaign including some measurement set-points where the chamber humidity and thermal imager angle were varied. The ILW container temperature was varied from \SIrange{15}{45}{\celsius}; the environmental chamber temperature from \SIrange{10}{25}{\celsius}; environmental chamber humidity at \SI{60}{\percent}rh and \SI{90}{\percent}rh; and the angle between thermal imager angle and vent normal at \SI{50}{\degree} and \SI{70}{\degree}.

\subsection{Phosphor thermometer}\label{subsec:measurements_phosphor}
The phosphor thermometer was positioned behind a set of lenses that allowed remote operation of the probe, the same phosphor coating as reported in~\cite{ref:ilw_paper_2018} was used. The probe consists of a combined illumination and measurement system that excites the phosphor coating with light and measures the decay time of phosphor emission. This instrument was traceably calibrated across the temperature range from \SIrange{1.6}{53.6}{\celsius} through a decay time comparison between a phosphor coated stainless-steel disc (\SI{25}{\milli\metre} diameter and \SI{5}{\milli\metre} thick) and an embedded calibrated N-type thermocouple below the coated surface.

\subsection{Thermal imager}\label{subsec:measurements_thermal_imager}
A long-wave infrared (LWIR) (\SIrange{7.5}{13.5}{\micro\metre}) FLIR Tau 2 microbolometer thermal imager was used. To minimise the effect of varying environmental temperatures on the instrument, it was mounted within a water-regulated brass enclosure (jacket) that was set to \SI{20}{\celsius}. 

The validation of apparent radiance temperature measured against ITS-90 was demonstrated through the calibration of the detector gain~\cite{ref:drift_correction} against a cavity reference source~\cite{ref:npl_cavity_references}. Whilst the instrument was mounted in its water-regulated enclosure, a two-point non-uniformity correction was measured at \SI{20}{\celsius} and \SI{30}{\celsius}, flooding the field of view; this narrow range was used to increase the responsivity to the anticipated application radiance levels. Through a measurement at \SI{23.5}{\celsius}, the small standard deviation of the digital level -- measured within an identical sized region of interest used within the vent measurements -- verified the higher image contrast than is typically achieved in the off-the-shelf configuration.

Following the non-uniformity correction, the instrument response was compared against the same blackbody reference source from \SIrange{5}{55}{\celsius} and a third-order polynomial used to describe the relationship between ITS-90 and detector digital level. The size-of-source effect of the instrument was characterised and the necessary correction from the calibration aperture to the size of the vent was applied throughout the vent measurements~\cite{ref:sse}.

As discussed in section~\ref{subsec:difference_results_technique}, the thermal imager measurements indicated that the calibration changed during the measurement campaign; so the thermal imager measured temperatures should be considered as indication of the relative temperature as opposed to an absolute temperature measurement.


\section{\bf Results} \label{sec:results}
This section details the results of the measurement undertaken within the environmental chamber. The results include both the thermal imager measurements and the phosphor thermometer measurements. The results cover measurements at a variety of chamber temperatures and humidities, as well as a range of ILW container temperatures. 

\subsection{Phosphor thermometer}
\begin{figure}[t]
\centering
\includegraphics[width=0.45\textwidth]{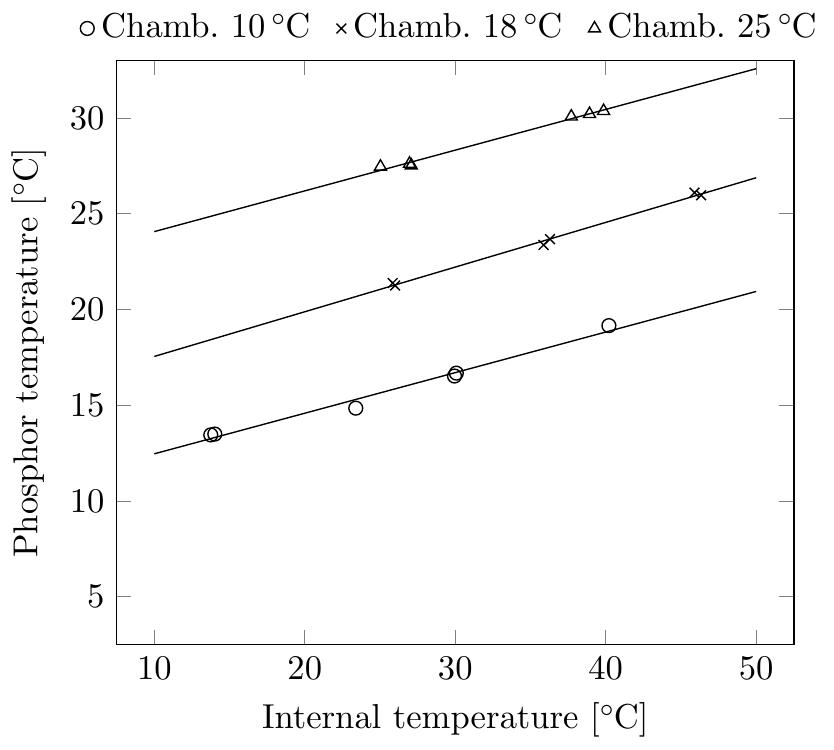}
\caption{The temperature of the vent measured by the phosphor thermometer plotted against the average container internal temperature, as measured by the PRTs. The points are differentiated by the three chamber set-point temperatures. The expanded measurement uncertainty of the phosphor thermometer measurements is \SI{0.11}{\celsius}.}
\label{fig:prt_v_phos_environ}
\end{figure}

Fig.~\ref{fig:prt_v_phos_environ} shows the temperature of the vent measured by the phosphor thermometer against the average container internal temperature, as measured by the PRTs.  The three chamber temperature data sets are distinctly stratified: the temperature measured by the phosphor thermometer is correlated with the chamber temperature.

\subsection{Thermal imager}
\begin{figure}[t]
\centering
\includegraphics[width=0.45\textwidth]{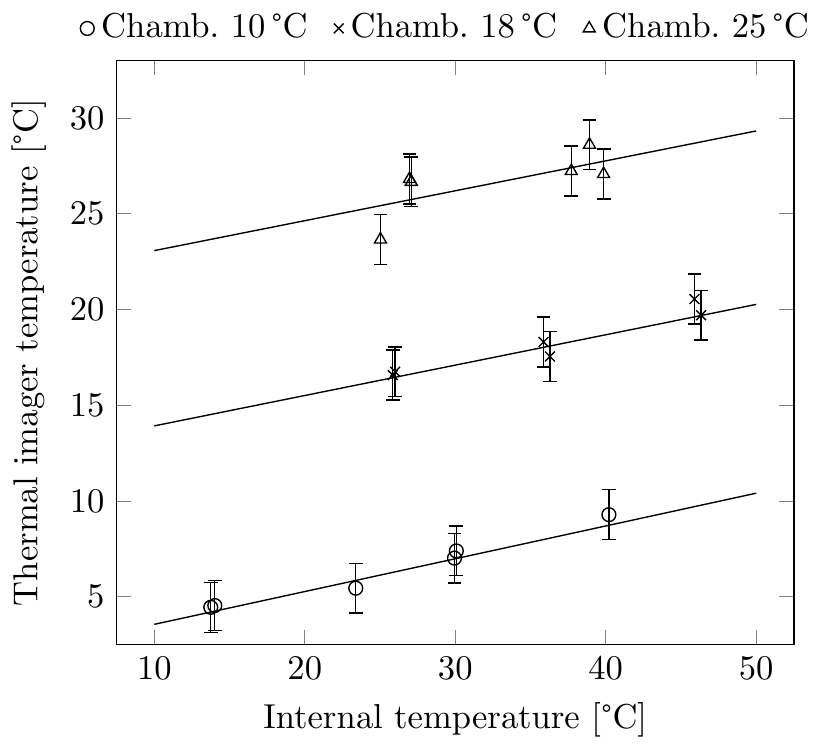}
\caption{The average temperature of the phosphor coated vents, as measured by the thermal imager, plotted against the average container internal temperature, as measured by the PRTs. The points are differentiated by the three chamber set-point temperatures. The error bars show the expanded measurement uncertainty.}
\label{fig:prt_v_ti_high}
\end{figure}

Fig.~\ref{fig:prt_v_ti_high} shows the average temperature of the phosphor-coated vents, as measured by the thermal imager, against the average container internal temperature, as measured by the PRTs. The linear fit is based on the average of temperatures measured for all phosphor coated vents per measurement set-point. Similar to the results recorded by the phosphor thermometer, the three chamber temperature data sets are distinctly stratified: the temperature measured by the thermal imager is correlated with the chamber temperature. It is of note that many of the data points in Fig.~\ref{fig:prt_v_ti_high} indicate that the measured temperature that is lower than the environmental chamber temperature (this is discussed in section~\ref{sec:discussion}). 

\subsection{Comparison between thermometry techniques}
\begin{figure}[t]
\centering
\includegraphics[width=0.45\textwidth]{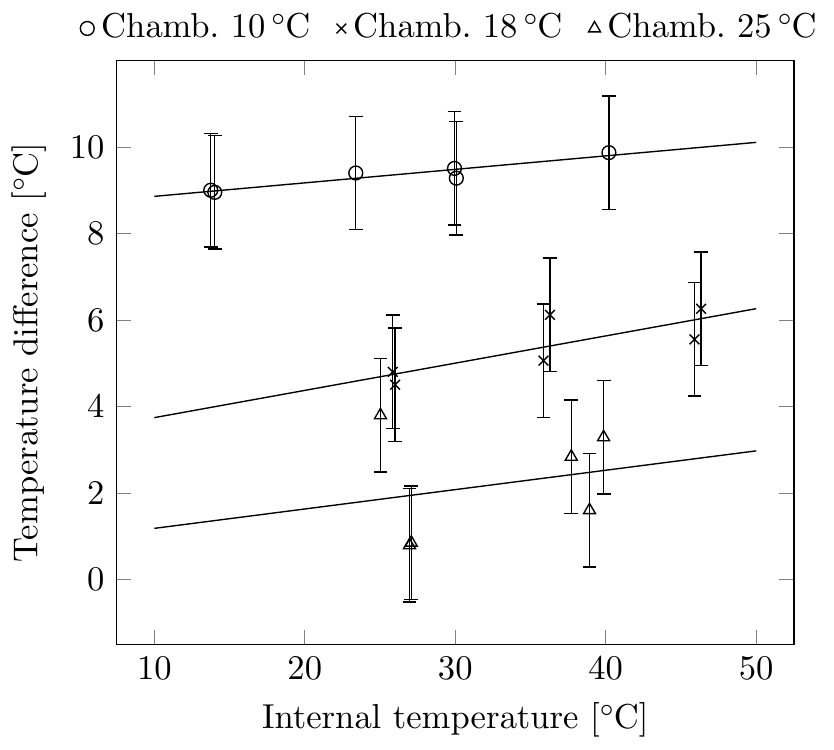}
\caption{The difference between the vent surface temperature measured by the phosphor thermometer ($T_{\mathrm{phosphor}} - T_{\mathrm{TI}}$), and the average surface temperature measurements of the phosphor coated vents, as measured by thermal imaging, plotted against  the temperature measured by the measured average of PRTs inside  the ILW container. The points are differentiated by the three chamber set-point temperatures.}
\label{fig:diff_prt_v_ti_phos}
\end{figure}

Fig.~\ref{fig:diff_prt_v_ti_phos} shows the difference between the vent surface temperature measured by the phosphor thermometer, and the average surface temperature measurements of the coated vents, as measured by thermal imaging. The measurement set-points with the largest difference in measured temperature occurred when the chamber was set to nominally \SI{10}{\celsius}. The set-points with the smallest temperature difference occurred when the chamber temperature was set to nominally \SI{25}{\celsius}. 

\section{\bf Uncertainty Budget} \label{sec:uncertainty_budget}
Following an analysis of the uncertainty components present for each of the three thermometry techniques (PRT, phosphor and thermal imaging), a budget was constructed for all measurements according to~\cite{ref:gum}. The majority of components considered were identical to those described in~\cite{ref:ilw_paper_2018}, the differences are detailed in this section.

\subsection{Contact thermometry}
The components evaluated were the same as those in~\cite{ref:ilw_paper_2018}. The expanded uncertainty of measurement for contact thermometry was \SI{0.27}{\celsius} \((k=2)\).

\subsection{Phosphor thermometry}
The components evaluated were the same as those in~\cite{ref:ilw_paper_2018}. The contact thermocouple used for the instrument calibration enabled a lower uncertainty to be achieved of \SI{0.06}{\celsius} \((k=2)\). Additionally, the standard deviation during measurement was less than \SI{0.05}{\celsius}. These two improvements in temperature metrology enabled an order of magnitude reduction of measurement uncertainty to \SI{0.11}{\celsius} \((k=2)\).

\subsection{Thermal imaging}
When considering the components used previously, this experimental setup has a  comparable expanded uncertainty of \SI{1.31}{\celsius} \((k=2)\). There were increases in the sensor stability and non-uniform emissivity components due to the varying environmental conditions, but these were dominated by the calibration component. It should be noted that the effect from non-unity emissivity has not been included within this budget.


\section{\bf Discussion} \label{sec:discussion}
This section discusses the results detailed in section~\ref{sec:results} and the uncertainty budgets detailed in section~\ref{sec:uncertainty_budget}. 

\subsection{Observations from results}
Figs.~\ref{fig:prt_v_phos_environ}~and~\ref{fig:prt_v_ti_high} show similar linear correlations. Both measurement techniques show the effect of the temperature from the local environment --- in this case, the environmental chamber --- on the surface temperature. The stratification of measurement results, dependent on chamber temperature, shows the effect of change in chamber temperature on the vent surface temperature; therefore, it is important that any measurements, whether they be recorded using thermal imaging or phosphor, are interpreted in reference to the local environmental temperature. 

The difference between the vent temperature measurements and the internal measurements indicates that, unlike in the previous results presented in~\cite{ref:ilw_paper_2018}, the vent temperature measurements, without taking environmental temperature into account, are a poor representation of the internal bulk temperature of the ILW container. 

As noted in section~\ref{sec:results}, Fig.~\ref{fig:prt_v_ti_high} presents measurements of the vent temperature that are lower than the environmental chamber temperature. For this experimental setup, it is not possible that the vent surface temperature was lower than the chamber temperature; this is due to the ILW container being fully situated within the chamber; this clearly indicates the presence of a systematic error in the thermal imaging measurements of the vent surface temperature. 

\begin{figure}[t]
\centering
\includegraphics[width=0.45\textwidth]{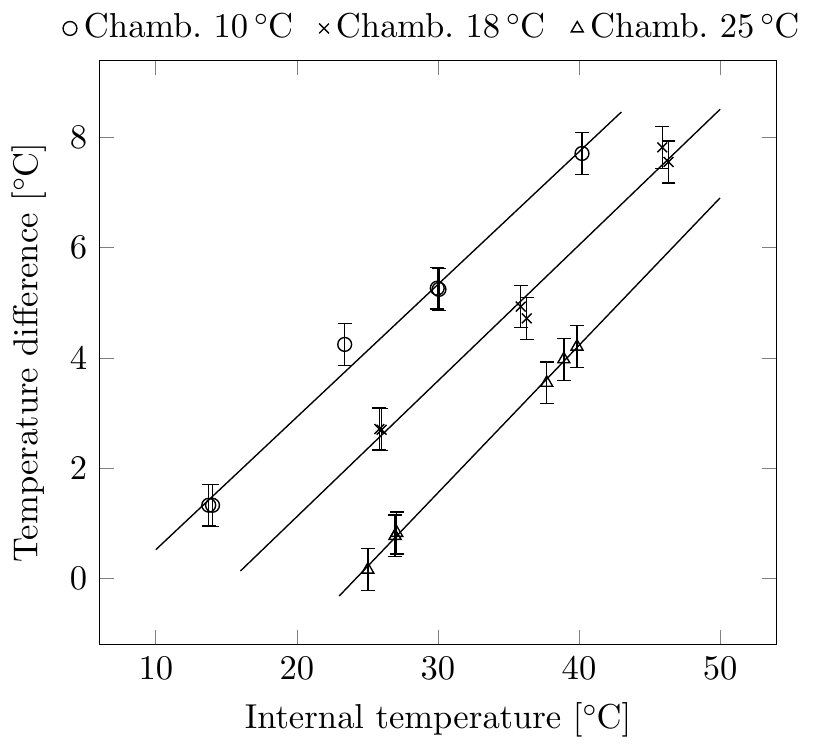}
\caption{The difference between the temperature measured by the PRT positioned at the bottom of the ILW container --- nearest the heat source --- and the PRT positioned at the top of the ILW container ($T_{\mathrm{PRT, bottom}} - T_{\mathrm{PRT, top}}$), plotted against the average temperature, as measured by all three PRTs.}
\label{fig:prt_top_bot_diff}
\end{figure}

Fig.~\ref{fig:prt_top_bot_diff} shows the difference between the temperature measured by the PRT positioned at the bottom of the ILW container --- nearest the heat source --- and the PRT positioned at the top of the ILW container, plotted against the average temperature, as measured by all three PRTs. The difference has greatest magnitude when the environmental chamber temperature was set to \SI{10}{\celsius}. The difference between the PRT measured temperatures is greatest when  the difference between internal container temperature and  the environmental temperature is greatest. This phenomena is a strong indication that the external surface temperature of the vent is not only dependent on the internal temperature of the container, but also the temperature gradient between the internal container temperature; as well as the external environment temperature and the coupling between the environment and the container surface. 

\subsection{Secondary influences on thermal imaging results}
Before addressing the systematic error in the thermal imaging, possible secondary effects on the temperatures measured by the thermal imager are considered, namely humidity and angle.

\subsubsection{Humidity}
Tab.~\ref{tab:hum_var} shows details of set-points that tested the sensitivity of thermal imager measurements to change in the relative humidity within the environmental chamber , RH$_{\mathrm{chamb}}$. The four pairs of measurement set-points are equivalent in terms of: nominal ILW container temperature, chamber temperature, and thermal imager angle. To test thermal imager measurement sensitivity to chamber relative humidity, the RH$_{\mathrm{chamb}}$ was varied from, nominally, \SI{60}{\percent}rh to,  nominally, \SI{90}{\percent}rh for each pair.  

\begin{table}[ht]
\centering 
\caption{The details of the four set-point pairs, where the humidity was changed. The thermal imager viewing angle was set to \SI{50}{\degree} for all measurement set-points. $T_{\mathrm{cont}}$ is the nominal internal container temperature, $T_{\mathrm{chamb}}$ is the nominal environmental chamber temperature, RH$_{\mathrm{chamb}}$ is the nominal environmental chamber humidity, and $T_{\mathrm{vent}}$ is the vent temperature as measured by the thermal imager.}
\begin{tabular}{c c c c}
\hline\hline 
$T_{\mathrm{cont}}$ [\SI{}{\celsius}] & $T_{\mathrm{chamb}}$ [\SI{}{\celsius}] &  RH$_{\mathrm{chamb}}$ [\SI{}{\percent}] & $T_{\mathrm{vent}}$ [\SI{}{\celsius}] \\ 
\hline 
15 & 10 & 60 & 4.45 \\ 
15 & 10 & 90 & 4.54 \\ 
25 & 25 & 60 & 26.82 \\ 
25 & 25 & 90 & 26.67 \\ 
35 & 10 & 60 & 7.02 \\ 
35 & 10 & 90 & 7.40 \\ 
45 & 25 & 60 & 27.23 \\ 
45 & 25 & 90 & 27.08 \\ 
\hline 
\end{tabular} 
\label{tab:hum_var}
\end{table}

The difference between the measured average vent temperatures for the first set-point pair shown in Tab.~\ref{tab:hum_var} is \SI{0.09}{\celsius}; this is significantly less than the uncertainty of the measurements. The equivalent difference for the other set-point pairs are: \SI{0.15}{\celsius}, \SI{0.38}{\celsius}, and \SI{0.16}{\celsius}; these are also within the uncertainty of the measurements. These results indicate that the changes in relative humidity of the environment had negligible influence on the temperatures measured by thermal imaging.

\subsubsection{Angle}
Tab.~\ref{tab:ang_var} shows details of set-points that tested the sensitivity of thermal imager measurements to change in the thermal imager angle relative to the vent surface. The three pairs of measurement set-points are equivalent in terms of: nominal ILW container temperature, chamber temperature, and chamber relative humidity. To test thermal imager measurement sensitivity to thermal imager angle, the angle was varied from, nominally, \SI{50}{\degree} to,  nominally, \SI{70}{\degree} for each pair.  

\begin{table}[ht]
\centering 
\caption{The details of the four set-point pairs, where the thermal imager viewing angle was changed. The environmental chamber was set to \SI{60}{\percent}rh for all measurement set-points. $T_{\mathrm{cont}}$ is the nominal internal container temperature, $T_{\mathrm{chamb}}$ is the nominal environmental chamber temperature, TI$_{\mathrm{angle}}$ is the thermal imager viewing angle, and $T_{\mathrm{vent}}$ is the vent temperature as measured by the thermal imager.}
\begin{tabular}{c c c c}
\hline\hline 
$T_{\mathrm{cont}}$ [\SI{}{\celsius}] & $T_{\mathrm{chamb}}$ [\SI{}{\celsius}] &  TI$_{\mathrm{angle}}$ [\SI{}{\degree}] & $T_{\mathrm{vent}}$ [\SI{}{\celsius}] \\ 
\hline 
25 & 18 & 50 & 16.58 \\ 
25 & 18 & 70 & 16.75 \\ 
35 & 18 & 50 & 18.31 \\ 
35 & 18 & 70 & 17.55 \\ 
45 & 18 & 50 & 20.55 \\ 
45 & 18 & 70 & 19.70 \\ 
\hline 
\end{tabular} 
\label{tab:ang_var}
\end{table}

The difference between the measured average vent temperatures for the first set-point pair shown in Tab.~\ref{tab:ang_var} is \SI{0.17}{\celsius}; this is significantly less than the uncertainty of the measurements. The equivalent difference for the other set-point pairs are: \SI{0.76}{\celsius} and \SI{0.85}{\celsius}; these are also within the uncertainty of the measurements. There is, therefore, no difference in these measured temperatures within the uncertainty of the measurements. A more general relationship between viewing angle and measured temperature has been established in~\cite{ref:ilw_paper_2018}.

\subsection{Difference in results per technique} \label{subsec:difference_results_technique}
Fig.~\ref{fig:diff_prt_v_ti_phos} shows the difference between the vent temperature measured by the phosphor thermometer, and the average vent temperature measurements of the coated vents, as measured by thermal imaging; this difference is significantly larger than the independent measurement uncertainty of each technique, detailed in section~\ref{sec:uncertainty_budget}; therefore, an investigation was undertaken to determine the source of this measurement discrepancy. The aspects that may have caused the large difference between thermal imaging radiance temperature and phosphor thermometer temperature were: the size-of-source effect of the vent diameter on the radiance temperature, insufficient decoupling of the thermal imager housing temperature from the environment, and the emissivity of the surface. These are described below and then an in-situ validation of thermal imager temperature using phosphor thermometer measured vent temperatures is proposed.

\subsubsection{Size-of-source effect}
The apparent temperature of a thermal radiation source is dependent on the apparent size of the thermal radiation source as viewed by a thermal radiation measuring device --- in this case a thermal imager --- this effect is known as the size-of-source effect (SSE)~\cite{ref:sse}. If the magnitude of this effect is well understood for a given measurement scenario, a correction for radiance temperature can be applied based on the apparent size of the radiation source. The calibration of the thermal imager utilised a \SI{40}{\milli\metre} diameter aperture. During the chamber measurements, the projected diameter of each vent was \SI{8}{\milli\metre}. The determination of the necessary correction suggests there is a \SI{2}{\celsius} offset in the thermal imager measured temperature that is nominally invariant with temperature. This correction does not account for the measured difference in temperatures measured by the two techniques as detailed in Fig.~\ref{fig:diff_prt_v_ti_phos}, which is as large as \SI{10}{\celsius}.

\subsubsection{Surface emissivity}
The thermal imager measures apparent radiance temperature, which does not account for the non-unity value of the surface emissivity. The difference between the apparent radiance and true surface temperature for a surface with emissivity less than \num{1} will vary as a function of the surface temperature, but it will also be equal to \SI{0}{\celsius} when the surface is in thermal equilibrium with the environment. Assuming the phosphor thermometer is representative of the surface temperature, Fig.~\ref{fig:diff_prt_v_ti_phos} does not describe this relationship as the fits show that the temperature difference does not equal \SI{0}{\celsius} when the chamber and container internal temperatures are equal. Therefore the source of the difference in Fig.~\ref{fig:diff_prt_v_ti_phos} is due to effects other than the emissivity (see Appendix for more detail).

\subsubsection{Thermal imager thermal-regulator}\label{subsec:error_id}
A laboratory based evaluation of the suitability of the enclosure used to regulate the thermal imager housing temperature was undertaken. During the environmental chamber measurements, the jacket was maintained at \SI{20}{\celsius}. In the laboratory evaluation the thermal imager was set to measure a reference blackbody source increasing from \SIrange{5}{55}{\celsius}. This was repeated three times with the jacket temperature set at \SI{18}{\celsius}, \SI{20}{\celsius}, and \SI{22}{\celsius}. The results show no observable dependence of the radiance temperature measured on the jacket temperature --- the difference was within the instrumentation measurement uncertainty.

\begin{figure}[t]
\centering
\includegraphics[width=0.45\textwidth]{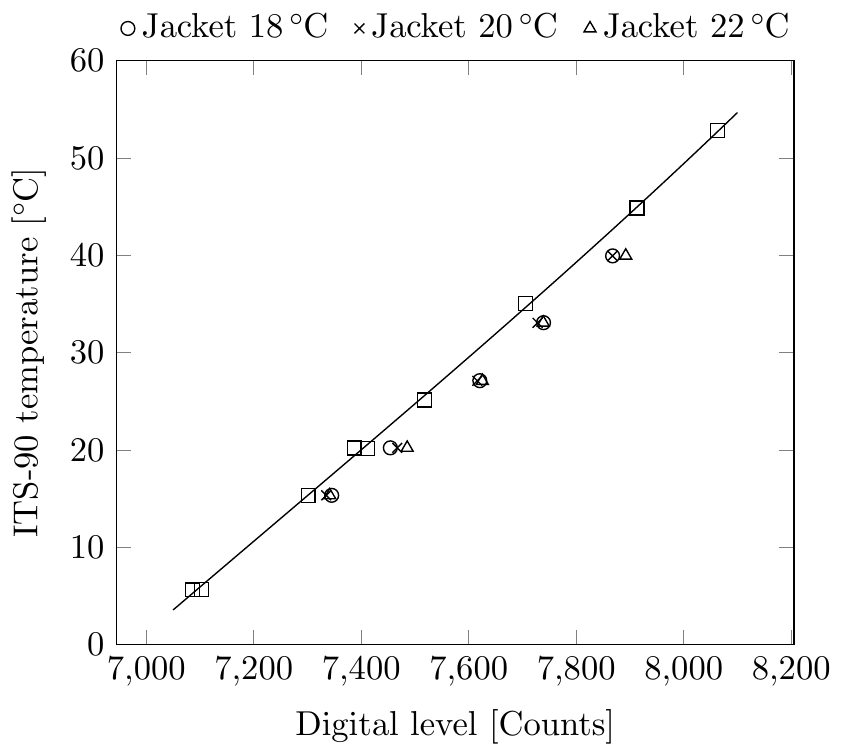}
\caption{Blackbody cavity temperature (ITS-90 temperature) plotted against the thermal imager focal plane array digital level --- the fit of the data original data set (\(\square\)) was used for all thermal imager ILW container vent measurements throughout this investigation --- plus the results of the additional measurements with varied jacket temperatures.}
\label{fig:ti_corrob_plus_mh_fpa}
\end{figure}

Fig.~\ref{fig:ti_corrob_plus_mh_fpa} shows the results of the correlation between digital level and temperature and a fit of results, used throughout this investigation to convert thermal imager outputted digital level to a corresponding surface temperature as described in section~\ref{subsec:measurements_thermal_imager}, as well as the three sets of additional measurements described. All three additional data sets are self-consistent, indicating that a change of jacket temperature within $\pm\SI{2}{\celsius}$ does not have a significant effect on digital level recorded.  

An offset between the original data set and the additional measurement results can be seen. It is of note that, between the original measurements and the additional measurements, the thermal imager was removed from the jacket and subsequently remounted in the jacket. It is understood that this remounting changed the thermal contact pathways between the imager and the jacket, thus changing the initial thermal dissipation characteristics. The effect of imager remounting within the jacket on the digital level recorded is important to the measurement of the ILW container vent temperature as the thermal imager was removed from the jacket after the initial calibration measurements and remounted in the jacket to measure the vent temperature in the environmental chamber. The effect of change in thermal contact between the thermal imager housing and an outside heat sink has been further investigated and shown to affect the focal plane array (FPA) digital level of  the thermal imager, as indicated by the results shown in Fig.~\ref{fig:ti_corrob_plus_mh_fpa}.

\subsubsection{In-situ calibration}\label{subsec:in-situ_cal}
Following the identification of the phenomenon described in section~\ref{subsec:error_id}, the viability of performing an in-situ calibration of the thermal imaging results using the results of the phosphor thermometry was investigated.

The in-situ calibration is based on a fit of the digital level and estimated FPA temperature with the phosphor measured vent temperature. The fit has functional form:

\begin{equation}
T_{\mathrm{phos}}(DL,T_{\mathrm{FPA}}) = \alpha + \beta \cdot  DL + \gamma \cdot T_{\mathrm{FPA}}~,\label{eq:phos_calib}
\end{equation}

\noindent
where $DL$ is digital level, $T_{\mathrm{FPA}}$ is the FPA temperature, $T_{\mathrm{phos}}$ is the vent  temperature measured by the phosphor thermometry. For the data set recorded when the environmental chamber was set to \SI{10}{\celsius}, $T_{\mathrm{FPA}}$ was assumed to be \SI{18}{\celsius}; for a chamber temperature of \SI{18}{\celsius}, the $T_{\mathrm{FPA}}$ was assumed to be \SI{19}{\celsius}; and for a chamber temperature of \SI{25}{\celsius}, the $T_{\mathrm{FPA}}$ was assumed to be \SI{20.5}{\celsius}. These FPA temperature assumptions are based on experience with the thermal imager in the specified environments.

Both $T_{\mathrm{phos}}$ and $DL$ are the average of measurements per experimental set-point. The values for $\alpha$, $\beta$, and $\gamma$ are  $\SI{-230.2}{\celsius}$, $\SI{0.037}{\celsius}$, and $-1.165$, respectively. The residual of the thermal imaging temperature resulting from the fit described in Eqn.~(\ref{eq:phos_calib}) from a one-to-one fit with the phosphor measured vent temperature is \SI{1.18}{\celsius}.

\subsubsection{Coupling to ILW container internal temperature}
\begin{figure}[t]
\centering
\includegraphics[width=0.45\textwidth]{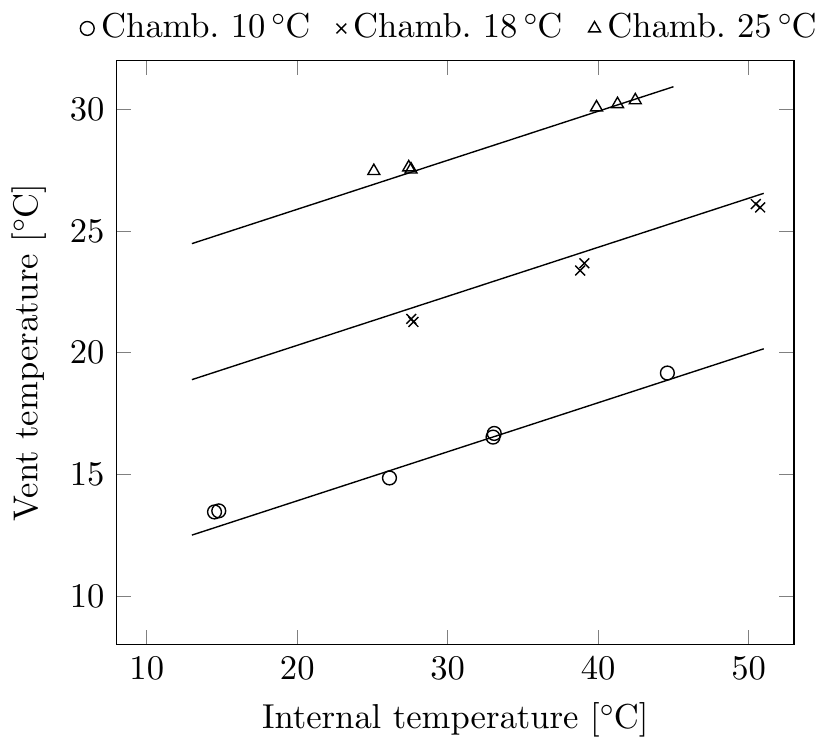}
\caption{The measured vent temperature plotted against the internal temperature, as measured by the PRT positioned at the bottom of the ILW container, for each of the environmental chamber set-points. The lines plotted describe the corresponding functions from Eqn. \ref{eq:container_temperature}.}
\label{fig:insitu_check}
\end{figure}

By using the in-situ calibration, it is possible to determine the coupling between the measured vent temperature and the internal temperature measured by the PRTs. Fig.~\ref{fig:insitu_check} shows the measured vent temperature plotted against the internal temperature, as measured by the PRT positioned at the bottom of the ILW container, for each of the environmental chamber temperature set-points. 

The fit shown in the figures was determined from the functional form:

\begin{equation}
T_{\mathrm{cont}} = T_{\mathrm{surf}} + \alpha(T_{\mathrm{surf}} - T_{\mathrm{chamb}}) + \beta~, \label{eq:container_temperature}
\end{equation}

\noindent
where $T_{\mathrm{cont}}$ is the PRT measured internal temperature, $T_{\mathrm{surf}}$ is the measured surface temperature of the vent, and $T_{\mathrm{chamb}}$ is the environmental chamber temperature. $\alpha$ and $\beta$ were determined to be $3.963$ and $\SI{-9.421}{\celsius}$, respectively. 

This plot presents the measured data from the experiment against Eqn. \ref{eq:container_temperature} and demonstrates the suitability of this function to determine internal container temperature. Whilst it is possible to infer the bulk internal temperature of the ILW container from the measured vent temperature for this experimental setup, the fit parameters will vary depending on the container geometry and material; the contents of the container; and the environmental conditions, in particular the environment temperature.


\section{\bf Conclusion} \label{sec:conclusion}
The results of the experiments carried out within the environmental chamber show the capability to measure the surface temperature of the vent on the ILW container lid and correlate this to the internal temperature of the container within a range of environmental conditions. It is particularly challenging to infer the internal temperature from the surface temperature for double skinned containers; and so approximating this property from the vent temperature provides insight to the rates of corrosion, hydrogen generation, thermal variations and radiogenic heating. However, to measure vent temperature reliably, the ambient temperature needs to be known and taken into account.

The results from both the phosphor thermometry and thermal imaging measurement techniques show that the vent temperature is sensitive to the temperature of the local environment. The stratification of measured temperature, dependent on the environmental chamber temperature, indicates as expected the vent temperature depends on the environment temperature. Further work would be required to generalise this work further, for example establishing the possible correlation between container surface and environment temperatures.

A significantly contributing source of the discrepancy between the measured vent temperature from the thermal imaging measurements and the phosphor thermometry measurements appears to have been a non-repeatable systematic error caused by the removing and remounting the thermal imager in a temperature controlled jacket. The non-repeatable thermal contact between the imager and the jacket has been shown to affect the outputted digital level whilst a stable scene is observed. Further work is required to determine the precise cause of this effect within the thermal imaging systems pipeline. 

By performing an in-situ calibration of the thermal imager measurements based on the phosphor thermometry, it has been shown to be possible to infer the bulk internal temperature of the ILW container from thermal imaging of the surface of the container vent. The value of using two independent surface thermometry techniques has been demonstrated by these measurements as solely the thermal imaging data would not have been suitable without either a correction from the phosphor thermometer or traceable surface emissivity data.

The uncertainty of thermal imaging measurements of the surface was approximately \SI{1.3}{\celsius}~(\(k=2\)). Measurements performed of the vent temperatures with varying relative humidity.  No significant change in vent temperature was measured indicating that the influence of relative humidity is negligible for the range of container and environment temperatures studied.


\section{\bf Acknowledgements} \label{sec:acknowledgements}
This research was funded through a commercial contract between NPL and Sellafield. The authors thank Sellafield for their continued support and collaboration.


\bibliographystyle{unsrt}
\bibliography{thermal_imaging_ilw_environmental_chamber}


\section{\bf Appendix} \label{sec:appendix}
\begin{figure}[ht!]
\centering
\includegraphics[width=0.45\textwidth]{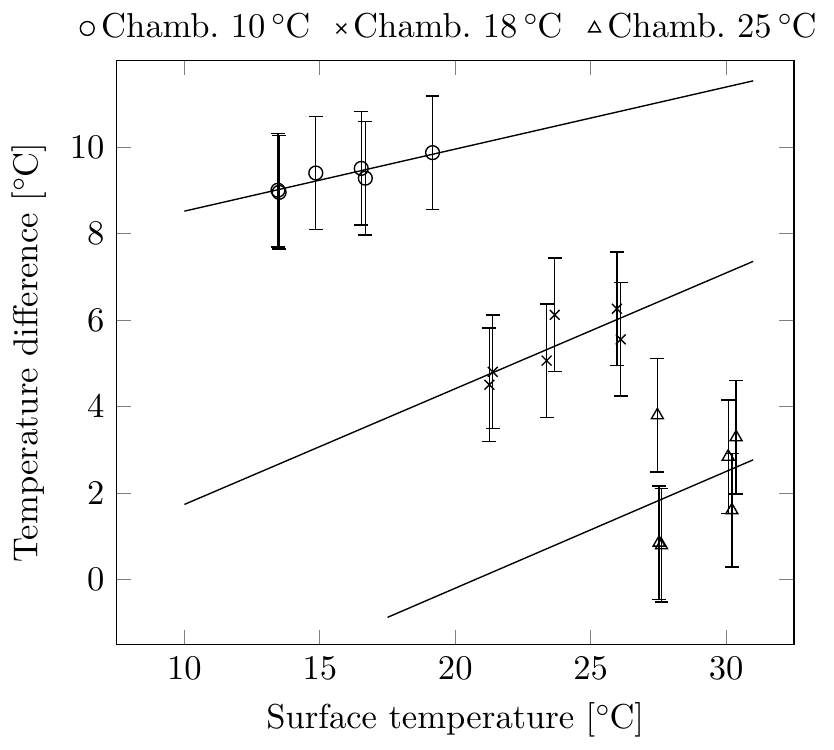}
\caption{An alternative representation of Fig.\ref{fig:diff_prt_v_ti_phos} as a function of the phosphor measured surface temperature. It is clear here that a zero temperature difference does not occur when the surface temperature and chamber temperature are equal.}
\label{fig:diff_phos_v_ti_phos}
\end{figure}

The data presented in Fig.~\ref{fig:diff_prt_v_ti_phos} is shown as a function of the internal PRT temperature. A representation of the same apparent radiance temperature difference instead as a function of the phosphor measured surface temperature can be seen in Fig.\ref{fig:diff_phos_v_ti_phos}. If the stratification was caused by an emissivity effect then the apparent radiance temperature difference would measure \SI{0}{\celsius} when the chamber and surface temperatures were equal.

\end{document}